\documentstyle[aps,epsfig]{revtex}
\newcommand{\nc}{\newcommand}
\newcommand{\sizeps}{10cm}
\nc{\rnc}{\renewcommand}
\rnc{\baselinestretch}{1.4}
\begin{document}
\begin{center}
{\Large \bf Weakly Nonextensive Thermostatistics and the Ising Model with 
Long--range Interactions} \\[5mm]  
{\bf  R. Salazar$^{1}$, A. R. Plastino$^{1,2,3}$, and  R. Toral$^{1,4}$}\\[4mm] 
{\small 
$^1$ Departament de F\'{\i}sica, Universitat de les Illes Balears, \\[-2mm]
07071 Palma de Mallorca, Spain.\\
$^2$ Faculty of Astronomy and Geophysics, National University La Plata, \\[-2mm]
Casilla de Correo 727, La Plata 1900, Argentina. \\
$^3$ CONICET, Casilla de Correo 727, La Plata 1900, Argentina. \\
$^4$ Instituto Mediterraneo de Estudios Avanzados (IMEDEA), \\[-2mm]
Campus UIB, 07071 Palma de Mallorca, Spain.
} \\[2mm] 
\end{center}

\newcommand{\be}{\begin{equation}}
\newcommand{\ee}{\end{equation}}
\newcommand{\ben}{\begin{eqnarray}}
\newcommand{\een}{\end{eqnarray}}
\newcommand{\pp}{\prime}
\newcommand{\nn}{\nonumber}

\newcommand{\ra}{\rangle}
\newcommand{\la}{\langle}
\newcommand{\ov}{\overline}

\vspace*{0.0cm}
\section*{Abstract}
We introduce a nonextensive entropic measure $S_{\chi}$ that grows like $N^{\chi}$,
where $N$ is the size of the system under consideration. This kind of nonextensivity 
arises in a natural way in some $N$-body systems endowed with long-range interactions 
described by $r^{-\alpha}$ interparticle potentials. The power law (weakly nonextensive)
behavior exhibited by $S_{\chi}$ is intermediate between (1) the linear (extensive) regime 
characterizing the standard Boltzmann-Gibbs entropy and the (2) the exponential law 
(strongly nonextensive) behavior associated with the Tsallis generalized $q$-entropies. 
The functional $S_{\chi} $ is parametrized by the real number $\chi \in[1,2]$ in such a 
way that the standard logarithmic entropy is recovered when $\chi=1$ . We study the 
mathematical properties of the new entropy, showing that the basic requirements for a well
behaved entropy functional are verified, i.e., $S_{\chi}$ possesses the usual properties
of positivity, equiprobability, concavity and irreversibility and verifies Khinchin axioms
except the one related to additivity since $S_{\chi}$ is nonextensive. For $1<\chi<2$, the 
entropy $S_{\chi}$ becomes superadditive in the thermodynamic limit. The present formalism
is illustrated by a numerical study of the thermodynamic scaling laws of a ferromagnetic 
Ising model with long-range interactions.
  
\vskip 0.5cm
\noindent
{\em Keywords:} entropy, nonextensivity, long-range interactions.

\noindent
{\em PACS:} 05.20.-y;05.20.Gg;02.70.Lq;75.10.HK;05.70.Ce
 
\newpage
\section*{1. Introduction}
\hspace*{\parindent} 
There is nowadays an intense research activity on the mathematical properties 
and physical applications of new versions of the Maximum Entropy Principle based 
on generalized or alternative entropic functionals
\cite{T88,ST99,PP97,M97,L99,LV98,A97,P98,BR98,AP99,RA99,RC99,F98,HGH98,Bo98,CJ96,B98,LMM98}.
This line of inquiry has been greatly stimulated by the work of Tsallis, who showed
that it is possible to build up a mathematically consistent and physically meaningful 
generalization of the standard Boltzmann-Gibbs-Jaynes thermostatistical formalism on 
the basis of a nonextensive entropic measure \cite{T88}. The main motivation
behind Tsallis proposal is that  there are important physical scenarios, such as 
self-gravitating systems \cite{S85,PP93b}, electron-plasma two dimensional
turbulence\cite{B95}, among many others, which are characterized by a nonextensive 
behavior: due to the long range of the relevant interactions some of the thermodynamic
variables usually regarded as additive, such as the internal energy, lose their extensive
character. This suggests that a {\em nonextensive} {\bf (non-additive)} entropy functional 
might be appropriate for their thermostatistical description. The Jaynes MaxEnt approach
to Statistical Mechanics \cite{jaynes,BS93} suggests in a natural way the possibility of 
incorporating \underline{ {\em alternative entropy functionals} }to the variational
principle. The new entropy functional introduced  by Tsallis has the form \cite{T88}

 \begin{equation} \label{tsaent}
 S_q=\frac 1{q -1}\left( 1-\sum_{i=1}^w p_i^q\ \right),
 \end{equation}

\noindent
where $(p_i, \,\,\, i=1, \ldots, w)$ are the microstate probabilities 
describing the system under consideration and the entropic index $q$ 
is any real number. The standard Boltzmann-Gibbs entropy 
$S = - \sum_{i=1}^w p_i \ln p_i $ is recovered in the limit $q\rightarrow 1$. 
The measure $S_q$ is
{\it nonextensive}: The entropy of a composite system $A \oplus B$ constituted
by two subsystems $A$ and $B$, independent in the sense that
$p^{(A \oplus B)}_{ij} = p^{(A)}_i p^{(B)}_j$, verifies the Tsallis' $q$-additive 
relation

\begin{equation} \label{qadi}
S_q(A \oplus B) = S_q(A) + S_q(B) + (1-q)S_q(A)S_q(B).
\end{equation}

\noindent
We see from the above equation that the Tsallis' parameter $q$ can be regarded as a 
measure of the degree of nonextensivity. Many relevant mathematical properties of the
standard thermostatistics are verified by the Tsallis' generalized formalism or can be 
appropriately  generalized \cite{T88,LMM98}.  
Self-gravitating systems constituted the first  physical problem discussed  within Tsallis'
nonextensive thermostatistics \cite{PP93b} and 
Tsallis' theory has recently been  applied to 
other physical problems \cite{TLSM95,LT98,BGM99,RML98,II98,II99}. 
 One of the main consequences of the intensive effort devoted in recent years to the study 
 of Tsallis  theory is that there is now a growing consensus that there are many 
 problems in statistical physics, biology, economics, and other areas, where a 
 generalization of the standard approach based on  Boltzmann-Gibbs-Jaynes extensive entropy
 might be useful. A comprehensive bibliography on the current research literature on Tsallis'
 theory and the statistical physics of nonextensive systems can be found in
 \cite{OBRAZIL,BRAZILEIRO}. 

Inspired on Tsallis' pioneering proposal, various nonextensive entropic 
measures endowed with interesting properties have been recently 
advanced \cite{LV98,A97,P98,BR98,AP99,RA99,RC99}. Moreover, it has been proved that some physically 
relevant mathematical properties are shared by large families of 
entropic measures \cite{PP97,M97}. The aim of the present work is to 
explore the possibility of developing a thermostatistical formalism based on a nonadditive
entropic functional characterized by a degree of nonextensivity weaker than the one 
exhibited by Tsallis measure. As we shall see, Tsallis entropy varies exponentially with the 
size of the system under consideration. The new measure introduced here 
only scales
as a power of the size of the system. That is to say, our proposal is tantamount of 
considering a nonextensive regime intermediate between (1) the standard extensive one 
associated with Boltzmann-Gibbs entropy, and (2) the exponentially nonextensive one 
described by Tsallis formalism. 

        There is an important physical motivation for introducing 
an entropy endowed with  power law nonextensivity. Systems with long range interactions
constitute the physical scenarios where the need of a generalization of the standard 
thermostatistical formalism can be more clearly appreciated. Consider a system of $N$ 
particles in a $d$-dimensional (one particle) configuration space. If the dependence of
the interparticle potential with the interparticle distance $r$ is given by $r^{-\alpha}$,
it can be shown that the system's energy levels scale as \cite{T95}

\be \label{nn*}
N \tilde N \, = \, N \,\, \frac{N^{1-\alpha/d} \, - \, \alpha/d}{1\, - \, \alpha/d}.
\ee
  
\noindent
Hence, for large $N$, the internal energy scales as a power of the size of the system. That is,

\be \label{enescal}
E \, \sim \, N^{2- \alpha/d}
\ee

\noindent
In the case of extensive systems governed by short-range interactions, 
the internal
energy $E$ and the the entropy $S$ scale in the same way: they both grow linearly
with $N$. On the other hand, the temperature $T$ is an intensive variable and does
not change with $N$. How can these scaling laws be generalized to the non-extensive
setting? A possible path towards the alluded generalization starts with the Helmholtz 
free energy,

\be \label{Helmholtz}
F \, = \, E \, - \, T S.
\ee

\noindent
From the above expression it seems reasonable to expect both $E$ and $T S$ to scale
in the same fashion \cite{T95}. The only way to fulfill this expectation, if the
standard extensive entropy is used, is to require that the temperature scales
as \cite{T95,CT96}

\be \label{tescal}
T \, \sim \tilde N \,  \sim  \, N^{1 - \alpha/d},
\ee

\noindent
losing its intensive character. 

It would be very appealing to have, within the nonextensive scenario, an entropy $S_\chi$
endowed with the same scaling law as the one exhibited by the energy. If, 
as it occurs for 
extensive systems, the entropy and the energy behave in the same way,  the temperature 
would still be an intensive quantity. What we need in order to have a thermostatistical
formalism with these nice properties is an  entropy functional with  power law
nonextensivity, scaling as

\be \label{sescal}
S_\chi \, \sim \, N^\chi \, \sim \, N^{2 - \alpha/d}. 
\ee

\noindent
From the above discussion it is clear that we are going to assume that
the exponent $\chi$ appearing in the entropic scaling law is related 
with $d$ and $\alpha $ by

\be \label{chiad}
\chi \, = \, 2 \, - \, \alpha/d, 
\ee

\noindent
so that the physically motivated range of vales for $\chi $ is

\be \label{chi12}
1 \, \le \, \chi \, \le 2.
\ee

The purpose of this work is to study the basic properties of a possible
candidate for a weakly-nonextensive thermostatistics, and to consider its 
application to a magnetic system with long-range interactions.
The paper is organized as follows. In Section II the exponential behavior 
of Tsallis entropy is analyzed and a power--law weakly nonextensive entropy 
is introduced. The basic mathematical properties of the new measure are studied
in Section III. Two state systems are considered in Section IV.  In Section V 
the weakly nonextensive entropy is applied to a ferromagnetic Ising model with 
long-range interactions. Finally, some conclusions are drawn in Section VI.

\section*{2. Exponential Vs. Power Law Nonextensivity}
%

\subsection*{Tsallis Entropy and  Exponential Nonextensivity}

    The nonextensive behavior of Tsallis functional, encapsulated in 
 equation (\ref{qadi}), is the most important single property distinguishing 
 Tsallis measure from the standard additive logarithmic entropy. An important
 consequence of relation (\ref{qadi}) that has not been fully appreciated in 
 the literature is that Tsallis entropy may varies exponentially with the size of 
 the system. In order to clarify the above assertion let us consider a system
 consisting of $N$ independent two-state subsystems. For simplicity we also assume
 that the system is described by an equiprobability distribution. That is, each
 of the $w=2^N$ possible microstate of the system has probability $1/w$. The 
 associated Tsallis measure adopts then the value
 $S_q^{equiprob.} \, = \, \frac{1 - w^{1-q}}{q-1}$,   
 so that, for $q<1$ and large $N$,

 \be \label{ndos}
 S_q \, \sim \,  \frac{1}{1-q} \,\,  2^{(1-q)N}.
 \ee

 \noindent
 If $q>1$ the $q$-entropy tends to the constant limit value $\frac{1}{q-1}$
 as $N\rightarrow \infty$. We are going to consider only the $q<1$ regime. 
 In that case the $q$-entropy exhibits an exponential behavior as a function of the size 
 $N$ of the system. In general, the entropy $S_q(N)$ of a composite system 
 $A=\oplus_{j=1}^N A^{(j)}$ consisting on $N$ identical and independent subsystems
 $\{A^{(j)}, \,\, j=1,\ldots N\}$, is given by
 
 \be \label{nrepli}
 S_q(N) \, = \, \frac{1}{q-1} \, 
 \left\{
 1 \, - \, [1+(1-q)S_q(1)]^N
 \right\},
 \ee
 
 \noindent
 where $S_q(1)$ stands for the individual entropy of each constituent subsystem $A^{(j)}$.
 Notice that in order to obtain the above expression it is not necessary to assume 
 that each subsystem $A^{(j)}$ is described by an equiprobability distribution.

\subsection*{Power Law Nonextensive Entropy}

We believe that it is worth exploring the possibility of a thermostatistical formalism
based on a nonextensive entropy $S_\chi$ growing as a power $N^\chi$ of the size of the 
system, instead of growing exponentially. In accord with the physical arguments discussed 
in the Introduction (see equations (\ref{chiad}-\ref{chi12})), we shall restrict our 
considerations to values of $\chi \in [1,2]$ . Let us assume that the functional $S_\chi$ 
associated with a given discrete probability distribution $\{p_i,\;i=1,\ldots,w\}$ 
is given by

\begin{equation} \label{schi}
S_\chi = \sum_{i=1}^w \phi_\chi(p_i),
\end{equation}

\noindent
$\phi_\chi(p_i)$ being an appropriate function of the individual microstate
probabilities $p_i$. In order to find a suitable expression for the function
$\phi_\chi(z)$ it is enough to consider again the equiprobability distribution
associated with a collection of $N$ identical independent two state subsystems.
 In that  case we have $w=2^N$ and

\ben \label{swchi}
S_\chi \, &=& \, w \, \phi_{\chi}(1/w) \cr
         &=& \, 2^N \phi_\chi(2^{-N}) \cr
	 &\sim & N^\chi, 
\een

\noindent
which implies

\be \label{swchi2}
\phi_\chi(2^{-N}) \, \sim \, 2^{-N} N^{\chi}.
\ee

\noindent
The simplest choice for a function $\phi_\chi (z)$ complying with the above 
relation is 

\be \label{simplest}
\phi_\chi(z) \, = \, z (-\ln z)^\chi.
\ee

\noindent
Unfortunately, this function is not adequate for our purposes. Its second derivative
$d^2\phi_\chi/dz^2$ lacks a definite sign within the relevant range of values of $z$ and
$\chi $, leading thus to an entropy functional without a definite concavity. 
However, since we are only interested in the large $N$ asymptotic regime,
any function $\phi_\chi (z)$ behaving like (\ref{simplest}) in the limit $z \rightarrow 0$ 
would do. As we shall see, the function

\be \label{schiz}
\phi_\chi(z) \, = \, \frac{1}{2} \, \left[ z (-\ln z)^\chi + z (-\ln z)^{1/\chi} \right]
\ee

\noindent
leads to the measure

\be  \label{schip}
S_\chi \, = \, \frac{1}{2} \, \sum_{i=1}^w \,
\left[ p_i (-\ln p_i)^\chi + p_i (-\ln p_i)^{1/\chi} \right],
\ee

\noindent
which complies with all the basic properties of a well behaved entropy. It
is clear that in the case $\chi=1$ the standard entropy
$S_1=-{\displaystyle \sum_{i=1}^w} p_i\ln{p_i}$  is recovered.

The optimization of $S_\chi$ under the constraints
imposed by normalization,

\begin{equation}
\sum_{i=1}^w p_i     \;=\;1  \label{constr1},
\end{equation} 

\noindent
and the mean values

\begin{equation}
\;\;\;\;\;\;\;\;\;\; \langle {\cal A}^{(r)} \rangle \;\equiv\;
\sum_{i=1}^w p_i {\cal A}^{(r)}_i \;=\;{\cal A}_\chi^{(r)} 
\;\;\;\;\;\;\;\;\;(r=1,\ldots,R)
\label{constr2}
\end{equation}

\noindent
of a given set $\{ {\cal A}^{(r)} \}$ of observable leads to the
variational problem

\begin{equation} \label{variat}
\delta \left\{
S_\chi \, - \, \sum_{r=1}^R\beta_r \langle {\cal A}^{(r)} \rangle
\, - \, \alpha \, \sum_{i=1}^w p_i 
\right\} \, = \, 0,
\end{equation}

\noindent
whose solution is of the form

\begin{equation}  \label{pechi}
p_i= F\left(\alpha +\sum_{r=1}^R\beta_r {\cal A}_i^{(r)} \right)
\;\;\;\;\;\;\; i=1,\dots,w. 
\end{equation}

\noindent
Here $\alpha$ and $\{ \beta_r \}$ are appropriate Lagrange multipliers
and $F(z)$ is the inverse function of 

\be \label{sder}
\phi_\chi^{\prime}(z) \, = \, \frac{1}{2}
\left\{
(-\ln z)^{\chi} +
(-\ln z)^{\frac{1}{\chi}} - 
\chi (-\ln z)^{\chi -1} -
\frac{1}{\chi}(-\ln z)^{\frac{1}{\chi}-1} 
\right\},
\ee

\noindent
which verifies the relation

\be \label{sinver}
F[\phi_\chi^{\prime}(z)] \, = \, \phi_\chi^{\prime}[F(z)] \, = \, z.
\ee

\begin{figure}
\centerline{
\epsfig{figure=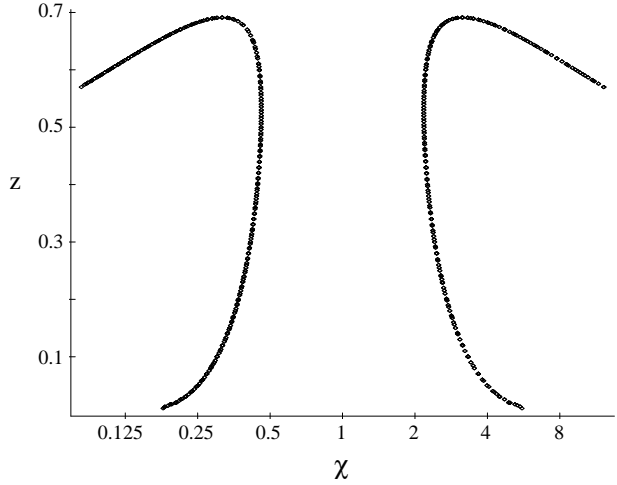,width=\sizeps}}
\caption[]{
The real roots of $s^{\prime \prime}_\chi(z)=0$ as a function of $\chi $.
There are no real roots within the interval $[1/\chi_c,\chi_c]$, where
$\chi_c \approx 2.1762$. 
\label{Fig1}}
\end{figure}

\noindent
The function $F(z)$ is well defined because the second derivative
of $s_\chi(z)$ is negative,

\ben \label{dsidpi}
\frac{{\rm d}^2s_\chi}{{\rm d} z^2}\; &=&\;
\frac{1}{2z} 
\left[
\chi (\chi -1) (-\ln z)^{\chi-2} \, +
\frac{1}{\chi} \left(\frac{1}{\chi} -1 \right)  (-\ln z)^{\frac{1}{\chi}-2} \,\right. \cr
&& \,- 
\left. \chi (-\ln z)^{\chi-1} \, +  \frac{1}{\chi} (-\ln z)^{\frac{1}{\chi}-1} \right] \cr
&<& 0, 
\een

\noindent
for $0<z<1$, and $1<\chi<2$. Actually, $s^{\prime \prime}_\chi(z)$ has a definite
(negative) sign within the larger interval $[1/\chi_c,\chi_c]$, where 
$\chi_c \approx 2.1762$. This
can be appreciated in Fig. \ref{Fig1}, where the real roots of 
$s^{\prime \prime}_\chi(z)=0 $ are depicted as a function of $\chi $. 
We see that there are no real roots $\chi \in [1/\chi_c,\chi_c]$. 

\begin{figure}
\centerline{
\epsfig{figure=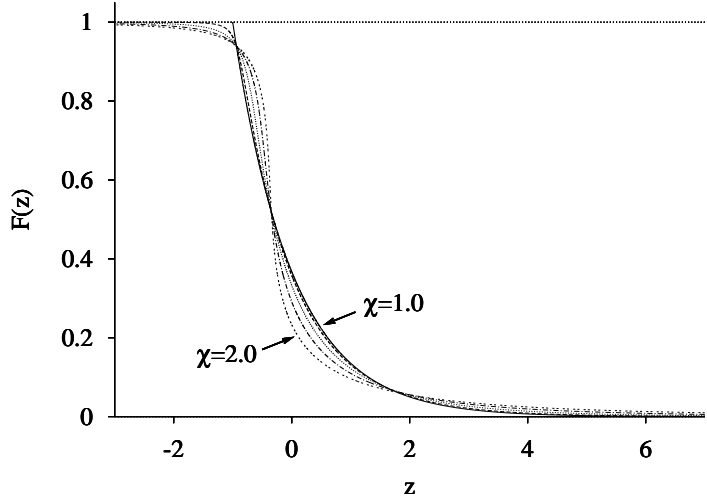,width=\sizeps}}
\vspace{0.5cm}
\caption[]{
The function $F(z)$ defined by the equations (\ref{sder}-\ref{sinver})
for $\chi=1.0,1.25,1.5,1.75,2.0$. 
\label{Fig2}}
\end{figure}

It is important to realize that the lack of a simple analytical expression for the 
function $F(z)$ does not constitute a serious conceptual problem, nor does it
pose any practical difficulty for the numerical treatment of problems involving the
MaxEnt distributions (\ref{pechi}). The form of the function $F(z)$ is shown in
Fig. \ref{Fig2}.

\vspace{1cm}

  The canonical ensemble probability distribution associated with the entropy 
  $S_\chi$,
  
  \begin{equation} \label{chicano}
  p_i \, = \, F(\alpha \, +\, \beta \epsilon_i), 
  \end{equation}
  
  \noindent
  is obtained when the $S_\chi $ is extremalized under the constraints 
  of normalization and the mean value of the energy,

  \begin{equation} \label{meane}
  \langle U \rangle \, = \, \sum_i p_i \epsilon_i,
  \end{equation}

  \noindent
  where $\epsilon_i $ stands for the energy of the microstate $i$. 
  The Lagrange multiplier $\beta$ appearing in (\ref{chicano}) is the
  one associated with the energy constraint and corresponds, within the
  present thermostatistical formalism, to the inverse temperature. That is,
  taking Boltzmann constant $k=1$, we have $\beta = 1/T$.  

\section*{3. Main Properties of the New Entropy}
\subsection*{3.1 Khinchin axioms}
\hspace*{\parindent} 
Khinchin proposed a set of four axioms\cite{khinchin}, which are
usually regarded  
as reasonable requirements for a well behaved information measure. 
Our entropy measure $S_\chi$ verifies the first three of them: \\[2mm]
(i)   $\;\;\;\;\;S_\chi=S_\chi(p_1,\ldots,p_w)$, \\
i.e., the entropy is a function of the probabilities $p_i$ only.
\\[2mm]
(ii)  $\;\;\;\;S_\chi(p_1,\ldots,p_w)\;\leq\;S_\chi(\frac{1}{w},
\ldots,\frac{1}{w}) \;\equiv\;S_\chi^{equipr.}(w)$, \\
i.e., $S_\chi$ adopts its extreme at equiprobability 
(this property will be proved in subsection 3.2). \\[2mm]
(iii) $\;\;\;S_\chi(p_1,\ldots,p_w)\;=\;S_\chi(p_1,\ldots,p_w,0)$, \\
this property, known as {\em expansibility}, is clearly verified since 
$\phi_\chi(0)=0$. \\[2mm]
(iv) $\;$ The fourth Khinchin axiom concerns the behavior of the entropy
of a composite system in connection to the entropies of the
subsystems. We will comment on this axiom later. 
\subsection*{3.2 General mathematical properties}
\hspace*{\parindent} 
Let us now consider the properties related to positivity, certainty,
concavity, equiprobability, additivity and irreversibility, which
are some of the most important features characterizing an information 
or entropic measure \cite{T88,P98,RC99,F98}.

\subsubsection*{3.2.1 Positivity}
It is plain from equation (\ref{schiz}) that $ \phi_{\chi}(0)=\phi_{\chi }(1)=0$ and also
that  $\phi_{\chi}(p) \ge 0$ for $p\in [0,1]$. This implies the {\em positivity condition} 

\begin{equation} \label{positiv}
S_{\chi }\geq 0.
\end{equation}
\subsubsection*{3.2.2 Certainty}
The equality symbol in Eq. (\ref{positiv}) holds only at {\em
certainty}, i.e.,  

\begin{equation} \label{border}
S_\chi(1,0,\ldots,0)\;=\;0.
\end{equation}
\\
Indeed, $S_\chi$ vanishes {\em if and only if} we have certainty. 
\subsubsection*{3.2.3 Concavity}
Considering $(p_1,\ldots,p_w)$ as independent variables, the
second partial derivatives of $S_\chi$ are 

\begin{equation}  \label{der2}
\frac{ \partial^2S_\chi}{\partial p_j \partial p_k}\;=\;
\frac{ \partial^2S_\chi}{\partial^2 p_j}  \,\delta_{jk}
\;<0
\;\;\;\;\;\;\;\;\;\mbox{for}\;\;\; 0<p_j<1.
\end{equation}
\\
Expression (\ref{der2}) guarantees definite {\em concavity} over probability 
space (see, for instance, \cite{BS93}).
\subsubsection*{3.2.4 Equiprobability}
The probability distribution that extremizes $S_\chi $ under the normalization 
constraint  is

\begin{equation}  \label{equipro}
p_i \, = \, F(\alpha) \, = \, 1/w
\end{equation}

\noindent
Therefore, since $S_\chi$ has negative concavity, it is {\em maximal at 
equiprobability}.  A well behaved entropy should also be, at equiprobability, a
monotonically increasing function of the number of states $w$. For large
$w$ we have $S_\chi(w) \sim (\ln w)^\chi$. Therefore, for large values of
$w$, $S_\chi^{equipr.}(w)$ {\em is an increasing function of} $w$.
\subsubsection*{3.2.5 Nonextensivity}
 The nonextensive behavior of the entropy $S\chi$ is determined by the relation of 
the entropy of a composite system with the individual entropies of its constituent 
subsystems. Let us consider systems $A$ and $B$ with associated probabilities 
$\{a_i,\;i=1,\ldots,w_A\}$ and $\{b_j,\;j=1,\ldots,w_B\}$, respectively. If systems 
$A$ and $B$ are independent, i.e., the composite system $A\oplus B$ has associated 
probabilities $\{a_i b_j;\;i=1,\ldots,w_A;\;j=1,\ldots,w_B\}$, then the entropy 
$S_\chi(A \oplus B)$ of the composite system minus the sum of the entropies
of its subsystems, 

\be \label{deltab}
\Delta S_\chi(A,B)\,\equiv\,S_\chi(A\oplus B)-S_\chi(A)-S_\chi(B)
\ee

\noindent
is the quantity characterizing the nonextensive features of the measure $S_\chi$.
When $\Delta S_{\chi}(A,B)>0$($\Delta S_{\chi}(A,B)<0$) we have superaditivity 
(subadditivity). From the examination of particular examples we conclude that 
$\Delta S_{\chi}(A,B)$ does not have always the same sign. However, the region 
of probability space where $\Delta S_{\chi}(A,B)$ is {\it positive} is much larger 
than the region where that quantity is negative. Furthermore, if we consider $N$ 
identical subsystems (instead of just two of them), the region of probability space 
corresponding to subadditive behavior tends to vanish as $N$ grows. consequently,
the entropy $S_\chi$ becomes superextensive in the thermodynamic limit. Particular 
examples illustrating these features of the measure $S_\chi$ are provided in 
Section IV. Moreover, in Section V we are going to consider the
nonextensivity of $S_\chi$ in connection with the thermodynamic properties
of an Ising model endowed with long--range interactions.

\subsubsection*{3.2.6 Irreversibility}
One of the most important roles played by entropic functionals
within theoretical physics is to characterize the ``arrow of
time''. When they verify an $H$-theorem, they provide a
quantitative measure of macroscopic irreversibility.  
We will now show, for some simple systems, that the present
measure $S_\chi$ satisfies an $H$-theorem,  
i.e., its time derivative has a definite sign. 

Let us calculate the time derivative of $S_\chi$

\begin{equation} \label{dsdt}
\frac{ {\rm d}S_\chi }{ {\rm d}t }\;=\;\sum_{i=1}^w  
\,\frac{{\rm d}p_i}{{\rm d}t} s^{\prime}_\chi (p_i),  
\end{equation}
\\
for a system whose probabilities  $p_i$ evolve according to the
master equation 

\begin{equation} \label{mastereq}
\frac{{\rm d}p_i}{{\rm d}t}\;=\;\sum_{j=1}^w [P_{ji}p_j-P_{ij}p_i],
\end{equation}
\\
where $P_{ij}$ is the transition probability per unit time
between microscopic configurations $i$ and $j$.  
Assuming a system with a uniform equilibrium distribution and detailed balance,
i.e., $P_{ij}=P_{ji}$, we obtain from (\ref{dsdt}) 

\begin{equation} \label{dsdtmod}
\frac{{\rm d}S_\chi}{{\rm d}t}\;=\;
\frac{1}{2}\sum_{i=1}^w \sum_{j=1}^w\,P_{ij}\,(p_i-p_j)
\biggl( s^{\prime}_\chi (p_j) \, - \,  s^{\prime}_\chi (p_i)  \biggr).
\end{equation}

\noindent
Since $s_{\chi}^{\prime \prime}(p_i) < 0$, the quantities $(p_i-p_j)$
and $( s^{\prime}_\chi(p_j) \, - \,  s^{\prime}_\chi(p_i) )$ have 
the same sign. Then we obtain

\begin{equation} 
\frac{{\rm d}S_\chi}{{\rm d}t}\;\geq\;0.
\end{equation}
\\
The equality holds for equiprobability, i.e., at equilibrium,
while in any other cases the entropy $S_\chi$ increases with
time. Therefore, $S_\chi$ exhibits {\em irreversibility}.  
\subsubsection*{3.2.7 Jaynes thermodynamic relations}
It is noteworthy that, within the present MaxEnt formalism, the
usual thermodynamical relations involving the entropy, the
relevant mean values, and the associated Lagrange multipliers,
i.e.,   
\begin{equation} \label{legendre}
\frac{\partial S_\chi}{\partial \langle {\cal A}^{(r)} \rangle }\;=\;\beta_r
\end{equation}
are verified.
Hence, our formalism exhibits the usual thermodynamical Legendre
transform structure.  
Actually, this property is verified by a wide family of entropy
functionals\cite{PP97,M97}. A particular important example of 
(\ref{legendre}) is furnished by the canonical ensemble thermodynamic
relation
\begin{equation}
\frac{\partial S_\chi}{\partial \langle U \rangle }\;=\;\beta\;=\;\frac 1 T. 
\end{equation}

\section*{4. Two-state systems}
\hspace*{\parindent} 
In order to illustrate some of the above properties, we consider 
a two-state system (with associated probabilities $\{p,1-p\}$). In this case, 
$S_\chi$ only depends on the variable $p$. In fact, from its
definition, we have   

\ben \label{seta22}
S_\chi(p) \, = \, \frac 1 2 && \left[
p (-\ln p)^\chi + p (-\ln p)^{\frac{1}{\chi}} \right. \cr
&&+ \,  \left. (1-p) (-\ln (1-p))^\chi +
(1-p) (-\ln (1-p))^{\frac{1}{\chi}}\right],
\een
\\

\begin{figure}
\centerline{
\epsfig{figure=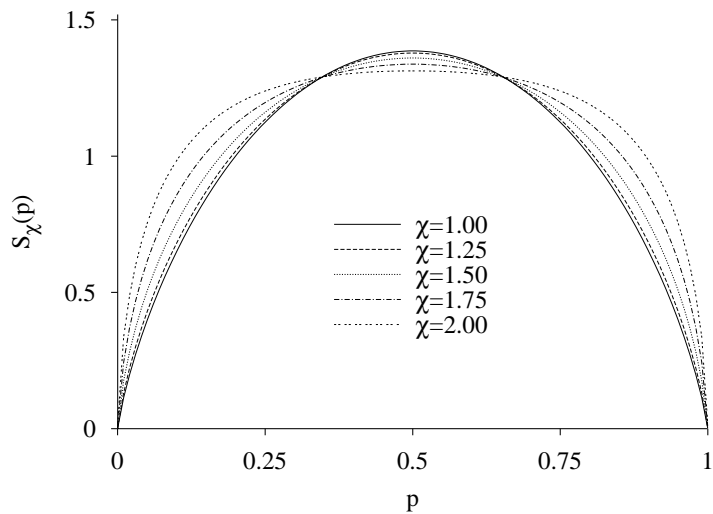,width=\sizeps}}
\vspace{0.5cm}
\caption[]{
 Entropy $S_\chi$ for a two-state system ($w=2$) as a function of
$p$ for different values of $\chi$ indicated on the figure. 
\label{Fig3}}
\end{figure}
\vspace{1cm}

The shape of $S_\chi(p)$ for different values of $\chi$ is
shown in Fig. \ref{Fig3}, which exhibits the positivity and
concavity of $S_\chi$.  
In fact, from expression (\ref{seta22}), the first derivative
of $S_\chi(p)$ vanishes at $p=1/2$ and ${\rm d}^2S_\chi/ {\rm
d}p^2 <0\;\;\forall p\in [0,1]$.
Since the second derivative is always negative,  $S_\chi(p)$ is 
maximal at equiprobability. Moreover, as shown in the general
case, taking into account the concavity of $S_\chi$ and that
$S_\chi$ vanishes at the certainty, then $S_\chi$ is positive for all $p$.  

\begin{figure}
\centerline{
\hspace{6cm}\epsfig{figure=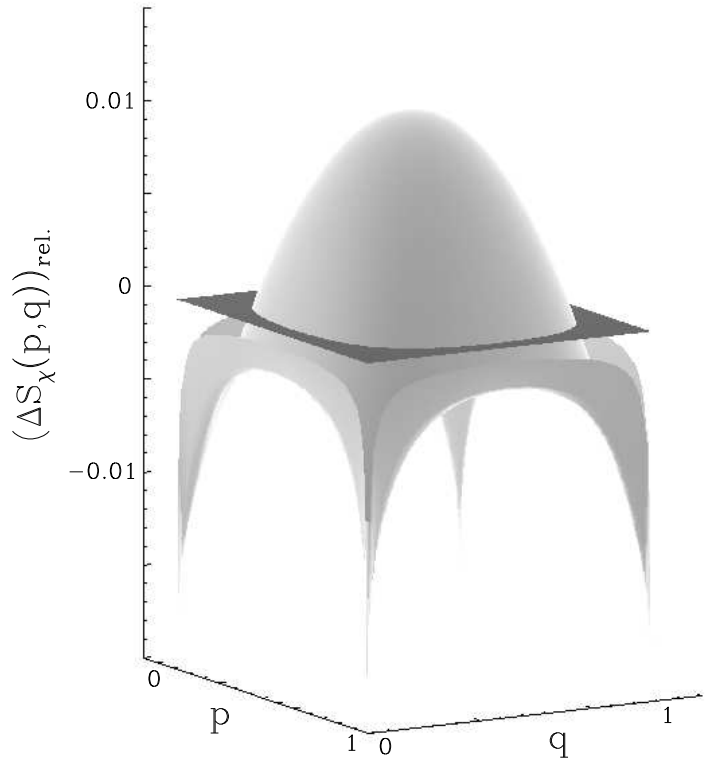,width=15cm}}
\vspace{1cm}
\caption[]{
The subadditivity and superaditivity of $S_\chi$. The quantity $(\Delta S_\chi)_{rel.}$ 
as a function of $p$ and $q$, for a composite system constituted by the independent 
 two-state subsystems $A$ (with associated probabilities $\{q,1-q\}$), and 
 $B$ (with associated probabilities $\{p,1-p\}$).
\label{Fig4}}
\end{figure}

The non-additivity of $S_\chi$ is illustrated in Fig. \ref{Fig4} for two independent 
two-state systems $A$ (with probabilities $p$ and $1-p$) and $B$ 
(with probabilities $q$ and $1-q$), through the plot of the relative difference 

\be \label{diffe}
(\Delta S_\chi)_{rel.} =[S_\chi(A\oplus B)-(S_\chi(A)+S_\chi(B))]/S_\chi(A\oplus B)
\ee

\noindent
as a function of $p$ and $q$. For most values of $p$ and $q$ the nonextensive measure 
$S_\chi$ behaves in a superadditive fashion. Only for values lying in a small region
near the edges of the $(p,q)-$square does $S_\chi $ become subextensive. 

\vspace{1cm}

  The nonextensive behavior of $S_\chi $ in the thermodynamic limit can be
  illustrate by recourse to a system constituted by $N$ two-state
  subsystem. Let us first assume that each one of the subsystems is 
  described by the same probabilities $p$  and $(1-p)$.  Consequently, the
  entropy  associated with the composite system is

  \be \label{enpep}
  S_\chi(N)  =  \frac{1}{2^N} \sum_{k=0}^N  
  \left(
  \begin{array}{c}
  N   \\  k 
  \end{array}
  \right)
  p^k  (1-p)^{N-k}
  \left[
  \left( -\ln( p^k \, (1- p)^{N-k} \right)^{\chi} +
  \left( -\ln( p^k \, (1- p)^{N-k} \right)^{\frac{1}{\chi}}
  \right]
  \ee

  \noindent
  It can be shown after some algebra that, for $0<p<1$,

  \be \label{chigrow}
  [-\ln(1-p)]^{\chi} N^{\chi} \, < \, S_\chi(N) \, < \, 
  [-\ln p]^{\chi} N^{\chi} + [-\ln p]^{1/\chi} N^{1/\chi}.
  \ee

\begin{figure}
\centerline{
\epsfig{figure=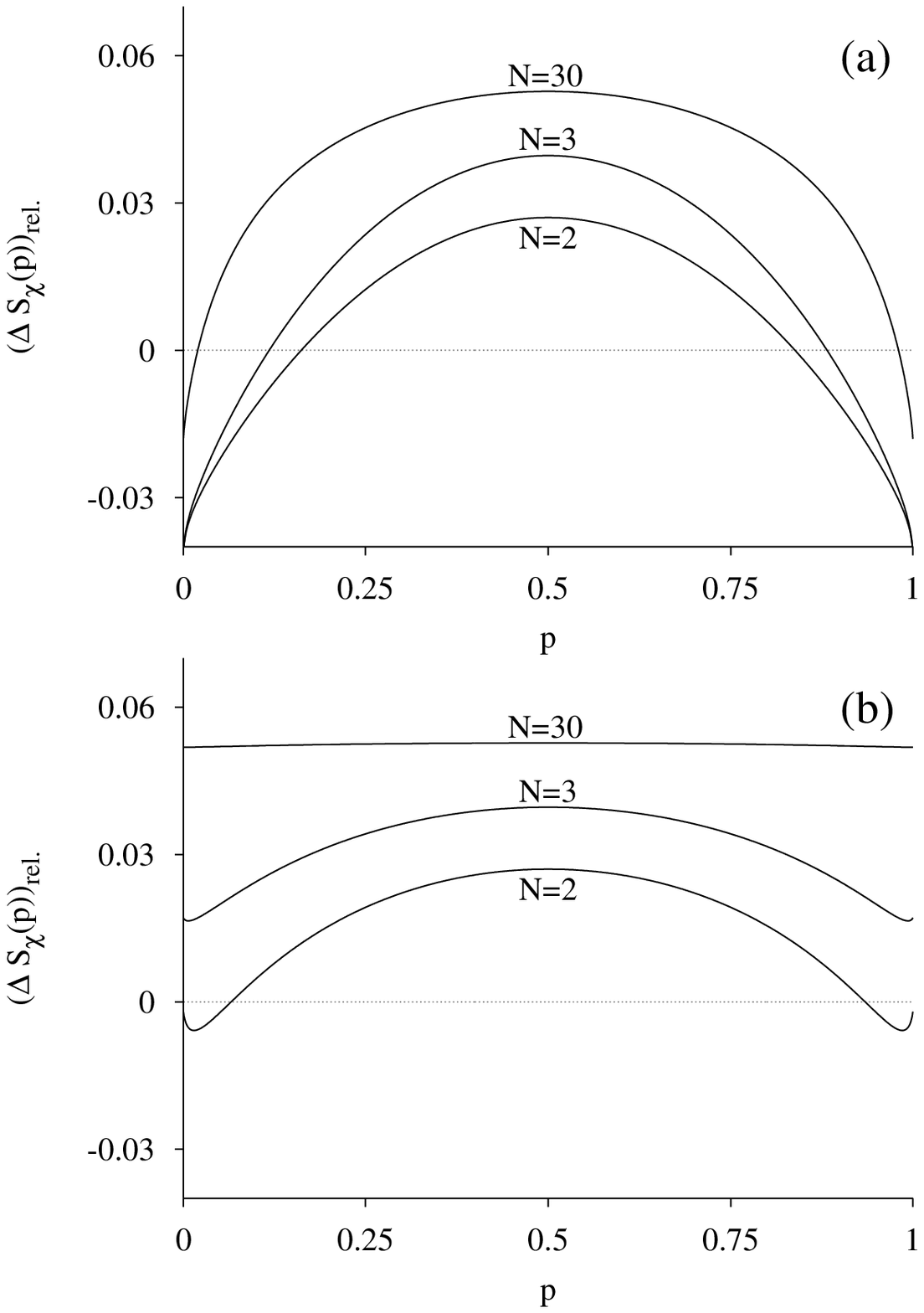,width=\sizeps}}
\caption[]{
 The  superaditivity of $S_\chi$ in the thermodynamic limit. 
(a) The quantity $(\Delta S_\chi)_{rel.}$ associated with a system constituted by $N$ 
identical two-state subsystems with probabilities $p$ and $(1-p)$, as a function of $p$
and for different values of $N$.
(b) $(\Delta S_\chi)_{rel.}$ as a function of $p$ for a system constituted
by $N-1$ identical two-state subsystems with probabilities
$\{ \frac{1}{2},\frac{1}{2} \}$ along with one extra two-state subsystem 
with probabilities $p$ and $(1-p)$.
\label{Fig5}}
\end{figure}

  \noindent
  Hence, for any given value of $p\in(0,1/2)$ there exist an $M$ such that

  \be \label{superad}
  N>M \, \Longrightarrow \, S_\chi(N) > N S_\chi(1),
  \ee

  \noindent
  which means that $S\chi $ becomes superadditive for large enough values of
  $N$. This is shown in Fig. \ref{Fig5}a , where the quantity

  \begin{equation}
  (\Delta S_\chi)_{rel.} \, = \, [S_\chi(N) \, - \, N S_\chi(1)]/S_\chi(N)
  \end{equation}

  \noindent
  is depicted for different values of $N$. In a similar way, Fig. \ref{Fig5}b shows
  the behavior of $(\Delta S_\chi)_{rel.}$ as a function of $p$ for a composite
  system consisting of $N-1$ two-state subsystems each with probabilities 
  $\{ \frac{1}{2},\frac{1}{2} \}$ along with one extra subsystem with probabilities
  $p$ and $(1-p)$.

\section*{5. Ferromagnetic Ising Model with Long-Range Interactions}

In this section we apply the canonical thermostatistics
associated with the entropy $S_\chi$ to an $N$-body system described by an
$r^{-\alpha}$ interparticle potential. The energy of this kind of systems scales 
as a power of the number of particles \cite{T95}. The consideration of systems exhibiting 
this power-law nonextensivity constituted our main motivation for introducing the 
measure $S_\chi$. As argued in Section I, an entropy endowed with the same non-extensive 
behavior as the one associated with the energy may lead to a thermostatistical 
description preserving the intensive character of the temperature. In order to
illustrate these ideas we are going to study a long-range Ising model described 
by the Hamiltonian

\be \label{Hamiltonian}
{\cal H} \, = \, J \, \sum_{i,j=1}^N \, \frac{1 \, - \, S_i S_j}{r^\alpha_{ij}},
\,\,\,\,\,\,\,\,\,\,\, (S_i = \pm 1, \forall i),
\ee

\noindent
where $J$ is an appropriate coupling constant, and the sum runs over all the pairs of 
sites on a $d$-dimensional lattice with periodic
boundary conditions and $r_{ij}$ stands for the distance between the sites $i$ and $j$. 
It is clear that the range of the interaction is determined by the value of the exponent
$\alpha$. In particular, the standard (short-range) first-neighbors interaction is 
recovered in the limit $\alpha \rightarrow \infty$, while the mean field approximation is 
obtained when $\alpha =0$ (replacing $J$ by $J/N$). These extreme cases illustrate, 
respectively, the two possible thermodynamic behaviors that, according to the value of 
$\alpha$, are admitted by the system (\ref{Hamiltonian}). On the one hand we have the 
extensive regime corresponding to $\alpha >d$. On the other hand we have the 
non-extensive regime associated with $\alpha <d$. In order to clarify this let us 
estimate the internal energy per particle at zero temperature. We have \cite{T95}

\begin{equation}
\frac{E}{N} \, \sim \, \int_1^{\infty} dr r^{d-1}r^{-\alpha}.
\end{equation}

\noindent
It is easy to see that the above integral converges if $\alpha>d$
and diverges if $0 \le \alpha \le d$. The following is a standard 
notation used in previous works:

\begin{equation}
\tilde N  \equiv 1+\frac 1 d \int_1^{N^{1/d}} dr r^{d-1}r^
{-\alpha}= \frac {N^{1-\alpha/d}-\alpha/d}{1-\alpha/d},
\end{equation}

Here we consider the non--extensive regime for the one--dimensional case
$d=1$ and $\alpha=0.8$. 
We have performed numerical simulations by using a novel approach 
recently introduced \cite{ST99} to study systems with long range interactions 
governed by generalized entropies as the one considered here. The 
method relies upon the calculation of the number of states, 
$\Omega(\epsilon_k)$, with a given energy $\epsilon_k$.

Note that, the number of possible configurations and
associated probabilities $p_i$ is in general very large, $W=2^N$ for Ising
models for example. However the number of permitted energies or energy levels $K$
is not so large, because there is a large number of states
$\Omega(\epsilon_k)$ sharing the same energy $\epsilon_k$. We can
rewrite the sums in Eqs. (\ref{schip},\ref{meane}) taking into account
the $K$ energy levels weighted by $\Omega(\epsilon_k)$:

\begin{eqnarray}\label{label1}
S_{\chi} &=& \frac 1 2 \sum_{k=1}^K \Omega(\epsilon_k) p(\epsilon_k)
[(-\ln(p(\epsilon_k))^{\chi}+(-\ln(p(\epsilon_k))^{1/\chi}], \\ 
<O> &=& \sum_{k=1}^K \Omega(\epsilon_k) p(\epsilon_k) O(\epsilon_k).
\end{eqnarray}
Hence, the knowledge of the number of states $\Omega(\epsilon_k)$ 
allows, by the use of these expressions, the calculation of the 
entropy and any averages of interest.

To calculate the probabilities $p(\epsilon_k)$ we use the definition
$p(\epsilon_k)=F(\alpha + \beta \epsilon_k)$, where the function $F$
is compute numerically as the inverse function of $s_{\chi}'(z)$ (See
Fig. \ref{Fig2}).

To compute the histogram $\Omega(\epsilon_k)$, the
``Histogram by Overlapping Windows (HOW)'' method
\cite{bha87,ST99,sal00} is used. A naive way of computing the 
histogram consists
in generating different system configurations randomly and counting how
many times a configuration with energy $\epsilon_k$ appears. However, 
since
the $\Omega(\epsilon_k)$ values span too many orders of magnitude it
is not possible to find in this way a histogram over all the energy levels.  The
HOW method avoids this problem by generating system configurations only
in a restricted energy interval and estimating the relative weights
$\Omega(\epsilon_k)/\Omega(\epsilon_l)$ of these energy levels
from the number of times they appear in the sample. From the overlap
between energy intervals, one gets the complete
$\Omega(\epsilon_k)$ function, apart from an irrelevant
normalization factor. Details about the HOW method can be found in
\cite{bha87,sal00}.

Finally we note a particularity of this new formalism, which is that
the determination of the $\alpha$ constant from the normalization
condition $\sum_i^w p_i=1$, require to solve the following equation

\begin{equation}
\sum_{k=1}^K \Omega(\epsilon_k) F(\alpha + \beta \epsilon_k)=1,
\end{equation}
where $\epsilon_k$,$\beta$,$\Omega(\epsilon_k)$ are input data for the
equation. 
This equation for $\alpha$ was solved numerically using a dicotomic
searching method.

Using the described procedure we have calculated the dependence
over the temperature $T$ of the internal energy $E(N,T)$, spontaneous
magnetization $M(N,T)=\mid \sum_{i=1}^N s_i \mid$, entropy $S(N,T)$
and free energy $F(N,T)=E-TS$ for the $1$--dimensional long--range
Ising model in the non--extensive regime $\alpha=0.8$ and the
corresponding value $\chi = 1.2$ in the Eq. (\ref{chiad}).

This system has been recently studied within the standard Boltzmann--Gibbs
thermostatistics \cite{CT96}, as well as within Tsallis non--extensive
$q$--formalism \cite{ST99}. In \cite{CT96} it was numerically verified that
the scaling laws of the main thermodynamical quantities associated with the
Gibbs canonical ensemble are

\ben \label{Gesca}
E(N,T)/(N\tilde N) \, &=& \, e(T/\tilde N), \cr
M(N,T)/N      \, &=& \, m(T/\tilde N), \cr
S(N,T)/N      \, &=& \, s(T/\tilde N), \cr
F(N,T)/(N\tilde N) \, &=& \, f(T/\tilde N),
\een

\noindent
Since the energy and the entropy scale in different ways, the temperature $T$
has to be scaled as $T/\tilde N$. A similar situation arises within Tsallis 
$q$-generalized formalism, the concomitant scaling laws being \cite{ST99} 

\ben \label{Tesca}
E_q(N,T)/(N\tilde N)   \, &=& \, e(T A^E_q(N) /(N \tilde N)), \cr
M_q(N,T)/N        \, &=& \, m(T A^E_q(N) /(N \tilde N)), \cr
S_q(N,T)/(A_q(N)) \, &=& \, s(T A^S_q(N) /(N \tilde N)), \cr
F_q(N,T)/(N\tilde N)   \, &=& \, f(T A_q(N) /(N \tilde N)),
\een

\noindent
where \cite{ST99}

\ben \label{qas}
A_q(N)   \, &=& \, (2^{N (1-q)}-1)/(1-q), \cr
A^S_q(N) \, &=& \, (2^{N |1-q|}-1)/|1-q|, \cr
A^E_q(N) \, &=& \, A_q(N)^2/ A^S_q(N).
\een

\noindent
The scaling laws corresponding to the generalized $\chi$-canonical
ensemble associated to the weakly nonextensive entropy $S_\chi$
are, 

\ben \label{Sesca1}
E(N,T)/(N\tilde N)                     \, &=& \, e(T R_\chi(N)), \cr
M(N,T)/N                          \, &=& \, m(T R_\chi(N)), \cr
S_\chi(N,T)/(N^\chi + N^{1/\chi}) \, &=& \, s(T R_\chi(N)), \cr
F(N,T)/(N\tilde N)                     \, &=& \, f(T R_\chi(N)),
\een

\noindent
where

\be  \label{rafescal}
R_\chi(N) \, = \, \frac{N^\chi \, + \, N^{1/\chi}}{N^{\chi} - (2-\chi)N}.
\ee

\noindent
It is plain from the above equation that

\be \label{easy}
\lim_{N \rightarrow \infty} R_\chi(N)  \, = \, 1,
\ee

\noindent
so that for large enough values of $N$ the scaling laws 
(\ref{Sesca1}) become

\ben \label{Sesca2}
E(N,T)/(N^\chi)      \, &=& \, e(T), \cr
M(N,T)/N             \, &=& \, m(T), \cr
S(N,T)_\chi/(N^\chi) \, &=& \, s(T), \cr
F(N,T)/(N^\chi)      \, &=& \, f(T).
\een

\begin{figure}
\centerline{
\epsfig{figure=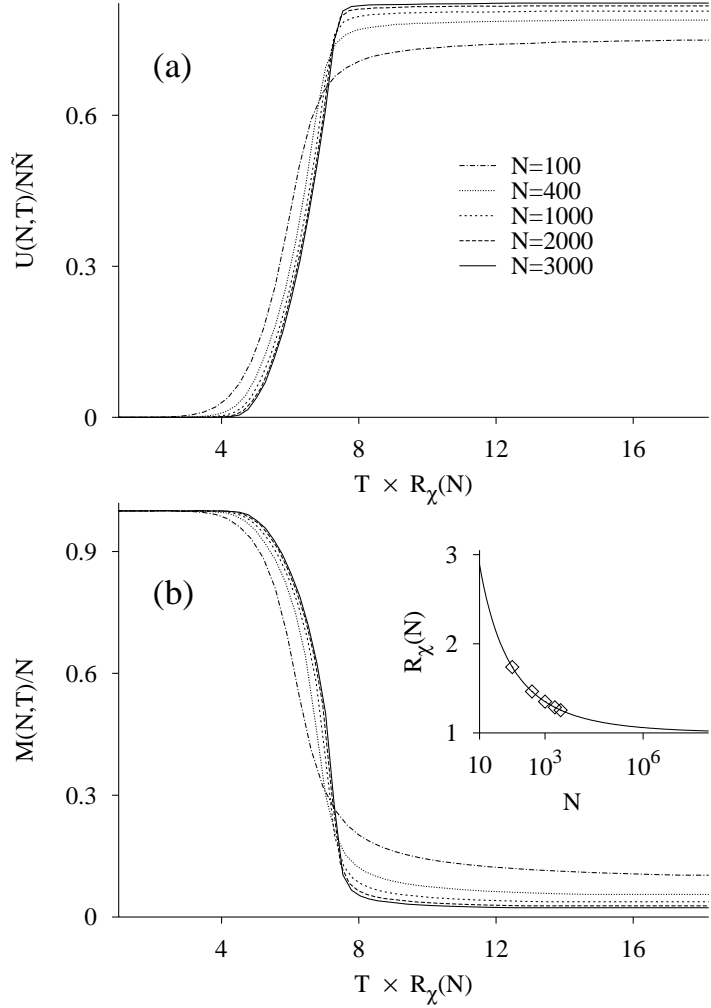,width=\sizeps}}
\vspace{0.5cm}
\caption[]{
Plot of (a) the internal energy and (b) the magnetization as a
function of the scaled temperature.  In the insert of (b) we plot with
solid line the behavior of the scaling factor $R_{\chi}(N)$ for all
the $N$ values and by points for the actual simulated sizes
$N=100,400,1000,2000,3000$. We use here $\chi=1.2$.
\label{Fig6}}
\end{figure}

\begin{figure}
\centerline{
\epsfig{figure=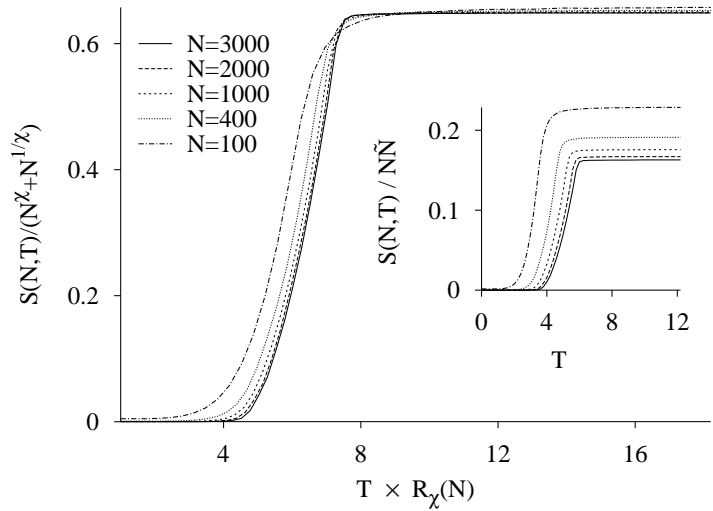,width=\sizeps}}
\vspace{0.5cm}
\caption[]{
Plot of the entropy $S_{\chi}$ as a function of the scaled temperature
for the indicated system sizes.  In the insert we plot the expected 
no scale temperature behavior.
\label{Fig7}}
\end{figure}

\noindent
According to the above equations, the thermodynamic curves of the Ising
models (\ref{Hamiltonian}) computed with increasing $N$-values must
collapse without the need of a temperature rescalation. Numerical evidence
of this scaling behavior is provided by Figures (\ref{Fig6},\ref{Fig7}). It
must be realized, however, that the $N$-values used are not large enough to
see a complete collapse of the curves depicted. In order to reach complete
collapse we need $R_{\chi}(N) \approx 1$ (see the insert of Figure
(\ref{Fig6}b)).

\vspace{1cm}

\noindent
We plot in Fig. \ref{Fig6} the scaled curves for the internal energy and
magnetization using the Eqs.  (\ref{Sesca1}). In the insert of that plot we
see the asimptotical behavior for the scaling factor $R_{\chi}(N)$, and with
points we indicate the sizes plotted $N=100,400,1000,2000,3000$. 
Unfortunately we would need very large sizes $N \sim 10^8$ to observe an
actual scaling without any scale factor, that is $R_{\chi} \sim 1$. However
from the Fig.6 one can see that there is an actual tendency to the
collapse. In fact we expect that the final universal curves will 
resemble to those corresponding to $N=2000,3000$ and we see in the Fig.
\ref{Fig6} a tendency towards a complete collapse.

\vspace{1cm}

Also in Fig.\ref{Fig7} we show the collapse for the entropy in the
main plot with the scaled relations Eqs. (\ref{Sesca1}) and in the
insert without scaling. In addition we can see in this plot the actual
superadditive behavior of the new entropy in an interacting system
instead of the simple $N$ independent two states systems of the last
section.

\section*{6. Conclusions}
\hspace*{\parindent} 
We have shown that the entropy functional $S_\chi$ verifies the
main properties usually regarded as essential requirements for
physically meaningful entropy functionals \cite{T88,P98,RC99,F98}. 
The entropy $S_\chi$ verifies the first three Khinchin axioms. 
The measure $S_\chi$ satisfies the requirements of positivity, equiprobability, 
concavity and irreversibility. The associated MaxEnt scheme also complies with 
Jaynes thermodynamic relations. The properties verified by $S_\chi$ suggest that 
it might be a useful measure for describing nonextensive  physical phenomena as well
as for other practical applications. 

The entropic functional $S_\chi$ grows as a power $N^{\chi}$ of the size $N$
of the system and is thus characterized by a weak nonextensive behavior (as
opposed to the strong, exponential nonextensivity exhibited by Tsallis
measure). An attentive reader might think that a power law nonextensive
measure can be obtained in a trivial way by just defining a new "entropy"
equal to $(S_{Boltzmann})^{\chi}$. However, this simple proposal would not
lead to anything new. The probability distributions maximizing 
$(S_{Boltzmann})^{\chi}$ are the same exponential distributions that
maximize $S_{Boltzmann}$. This means that the standard (extensive)
thermostatistics would be obtained again. On the contrary, the entropic
functional $S_{\chi}$ furnishes a nontrivial new thermostatistics that
might be appropriate for dealing with some nonextensive systems.
 As an application of the thermostatistical formalism associated with $S_\chi$
 we considered a ferromagnetic Ising model with long-range interactions. We
 studied numerically the concomitant scaling laws.
 
Tsallis thermostatistics, which have attracted a great deal of attention recently
\cite{TLSM95,LT98,BGM99,RML98,II98,II99,OBRAZIL,BRAZILEIRO}, arise from
a nonextensive entropic measure $S_q$ {\it growing exponentially with the size 
of the system}. Now, if there are in Nature extensive systems (that is, those
successfully studied within standard statistical mechanics and described by
Gibbs exponential distributions) as well as Tsallis' strong (exponentially)
nonextensive systems, it is not unreasonable to expect also the existence of
systems exhibiting an intermediate (weak) power-law kind of nonextensivity.

Let us consider a family of dynamical systems characterized by a set of
parameters $\lambda$. We assume that there is a certain region $\Sigma$ in 
$\lambda$-space such that the system's thermodynamic behavior is extensive
for $\lambda \in \Sigma$ and nonextensive otherwise. It would be very
surprising if, as a continuous change in the parameters $\lambda$ is
considered, the system abruptly jumps from an extensive regime to an
exponentially nonextensive one. It would be physically more sensible if
there is an "onset of nonextensivity" boundary between the extensive and
nonextensive regions in $\lambda$-space, characterized by a weak, power-law
nonextensive regime. A suggestive analogy can be made here with the onset of
chaos is nonlinear dynamics. Consider, for instance, the well known Logistic
map. For values of the map's parameter corresponding to chaotic behavior the
concomitant Lyapunov exponent is positive and near trajectories diverge
exponentially in time. However, critical values of the map parameter
associated with a vanishing Lyapunov exponent still do exhibit a form of
"weak chaos". In those critical situations, corresponding to the onset
chaos, near trajectories do not diverge exponentially in time: They diverge
in a power-law way \cite{TPZ97,CLPT97}.

It would be important to gain a further clarification of the physical meaning of the 
parameter $\chi $, and to understand under what circumstances might Nature
choose to maximize the functional $S_{\chi} $. The only way to attain such
an understanding is by a detailed study of the dynamics of particular nonextensive
systems endowed with long range interactions . We hope that our present 
contribution may stimulate further work within this line of inquiry.

\subsubsection*{Acknowledgements}
\hspace*{\parindent}
A.R.P. acknowledges the AECI Scientific Cooperation Program (Spain), 
and the CONICET (Argentine Agency) for partial financial support.
R. Toral and R. Salazar acknowledges the financial support from DGES, 
grants PB94-1167 and PB97-0141-C02-01.


\newpage
\centerline {\bf REFERENCES}

\begin{thebibliography}{99}

%
\bibitem{T88} C. Tsallis, {\em J. Stat. Phys.}  {\bf 52} (1988) 479;
C. Tsallis, R.S. Mendes, and A.R. Plastino, {\em Physica} {\bf A 261} 
(1998) 534; E.M.F. Curado and C. Tsallis, {\em J. Phys.} {\bf A 24} (1991) L69.

%
\bibitem{ST99} R. Salazar and R. Toral,  {\em Phys. Rev. Lett.} {\bf 83} (1999) 4233.

%
\bibitem{PP97} A. Plastino and A.R. Plastino, {\em Phys. Lett. } 
{\bf A 226} (1997), 257

%
\bibitem{M97} R. S. Mendes, {\em Physica } {\bf A 242} (1997), 299;
E.M.F. Curado, {\em Braz. J. Phys.} {\bf 29} (1999) 36.

%
\bibitem{L99} P.T. Landsberg,  {\em Braz. J. Phys.} {\bf 29} (1999) 46.

%
\bibitem{LV98} P.T. Landsberg and V. Vedral, {\em Phys. Lett.} 
{\bf A 247} (1998) 211.

%
\bibitem{A97} S. Abe, {\em Phys. Lett.} {\bf A 224}, 326 (1997).

%
\bibitem{P98} A.R.R. Papa {\em J. Phys.} {\bf A 31}, 1 (1998).

%
\bibitem{BR98} E.P. Borges and I. Roditi, {\em Phys. Lett.} {\bf A 246}, 399 (1998). 

%
\bibitem{AP99} C. Anteneodo and A.R. Plastino, {\em J. Phys. } {\bf A 32}, 1089  (1999).

%
\bibitem{RA99} A.K. Rajagopal and S. Abe, {\em Phys. Rev. Lett.} {\bf 83} (1999) 1711.

%
\bibitem{RC99} R. Rossignoli and N. Canosa {\em Phys. Lett.} {\bf A 264} (1999) 148.

%
\bibitem{F98} K.S. Fa, {\em J. Phys.} {\bf A 31}, 8159 (1998).  

%
\bibitem{HGH98} D. Holste, I. Grosse, and H. Herzel, {\em J. Phys.} {\bf A 31}, 2551 (1998).

%
\bibitem{Bo98} E.P. Borges, {\em J. Phys.} {\bf A 31}, 5281 (1998); 
E.K. Lenzi, E.P. Borges, and R.S. Mendes, {\em J. Phys.} {\bf A 32}, 8551 (1999).

%
\bibitem{CJ96} A. Compte and D. Jou, {\em J. Phys.}  {\bf A 29} (1996) 4321;
A. Compte, D. Jou, and Y. Katayama, {\em J. Phys.}  {\bf A 30} (1997) 1023.

%
\bibitem{B98} L. Borland, {\em Phys. Rev.} {\bf E 57} (1998) 6634.

%
\bibitem{LMM98} E.K. Lenzi, L.C. Malacarne and R.S. Mendes, {\em Phys. Rev.
Lett.} {\bf 80} (1998) 218.

%
\bibitem{S85}  W.C. Saslaw, {\em Gravitational Physics of Stellar and Galactic
                Systems}, Cambridge University Press, Cambridge, 1985.     

%
\bibitem{PP93b} A. R. Plastino and A. Plastino, {\em Phys. Lett.}  {\bf A 174} (1993) 384.


%
\bibitem{B95} B.M.R. Boghosian, {\em Phys. Rev.}  {\bf E 53} (1995) 4754.


%
\bibitem{jaynes} E.T. Jaynes, in {\em Statistical Physics}, ed.
                 W.K. Ford (Benjamin, New York, 1963); 
                 A. Katz, {\em Statistical Mechanics}, (Freeman,
                 San Francisco, 1967). 

\bibitem{BS93} C. Beck and F. Schl\"ogl, {\it Thermodynamics of Chaotic Systems:
               An Introduction} (Cambridge University Press, Cambridge, 1993).

\bibitem{TLSM95}  C. Tsallis, S.V.F. Levy, A.M.C. Souza and R. Maynard,
{\em Phys. Rev. Lett.} {\bf 75} (1995) 3589; Erratum: {\bf 77} (1996) 5442.

\bibitem{LT98} M. L. Lyra and C. Tsallis, Phys. Rev. Lett.  {\bf 80} (1998) 53;
C. Anteneodo and C. Tsallis, {\em Phys. Rev. Lett.}  {\bf 80}  (1998) 5313.

\bibitem{BGM99} M. Buiatti, P. Grigolini, and A. Montagnini, {\em Phys. Rev.
Lett.} {\bf 82} (1999) 3383.

\bibitem{RML98} A.K. Rajagopal, R.S. Mendes, and E.K. Lenzi, Phys. Rev.
Lett. 80 (1998) 3907.

\bibitem{II98} D.B. Ion and M.L.D. Ion, Phys. Rev. Lett.  81 (1998) 5714.

\bibitem{II99} M.L.D. Ion and D.B. Ion, Phys. Rev. Lett.  83 (1999) 463.

 \bibitem{OBRAZIL}C. Tsallis, {\em Physics World}  {\bf 10}, 42 (July 97);
an updated bibliography on Tsallis' theory an its applications
is available at http://tsallis.cat.cbpf.br/biblio.htm

\bibitem{BRAZILEIRO} Several review articles on Tsallis' 
thermostatistics and its applications appeared in {\em Braz. J. Phys.} {\bf 29} (1999), 
Special Issue, {\it Nonextensive Statistical Mechanics and 
Thermodynamics}, 
eds. S.R.A. Salinas and C. Tsallis, which can be seen at
http://sbf.if.usp.br/WWW{\_}pages/Journals/BJP/Vol29/Num1/index.htm

%
\bibitem{T95} C. Tsallis, {\em Fractals} {\bf 3}, 541 (1995).

%
\bibitem{CT96} S.A. Cannas and F.A. Tamarit, {\em Phys. Rev.} {\bf B54}, R12661 (1996).

%
\bibitem{khinchin} A.I. Khinchin, {\em Mathematical Foundations
                   of Information Theory} (Dover, New York, 1957).

\bibitem{bha87}
G.~Bhanot, R.~Salvador, S.~Black, P.~Carter, and R.~Toral.
\newblock {\em Phys. Rev. Lett.}, {\bf 59} (1987) 803.

\bibitem{sal00}
R.~Salazar, and R.~Toral.
\newblock  {\it Thermostatistics of extensive and Nonextensive
Systems Using Generalized Entropies}, (2000), preprint.

\bibitem{TPZ97}
C.~Tsallis, A.R.~Plastino, and W.-M.~Zheng.
\newblock {\em Chaos, Solitons and Fractals}, {\bf 8} (1997) 885.

\bibitem{CLPT97}
U.M.S.~Costa, M.L.~Lyra, A.R.~Plastino and C.~Tsallis.
\newblock {\em Phys. Rev. E}, {\bf 56} (1997) 245.

\end{thebibliography}
\end{document}